\documentclass{iaus}
\usepackage{graphicx, natbib, amsmath, multicol}

\newcommand{\aap}{\textit{A\&A}}
\newcommand{\araa}{\textit{ARA\&A}}
\newcommand{\apj}{\textit{ApJ}}
\newcommand{\apjl}{\textit{ApJL}}
\newcommand{\apjs}{\textit{ApJS}}

\newcommand{\mnras}{\textit{MNRAS}}
\newcommand{\nat}{\textit{Nature}}
\newcommand{\jcam}{\textit{J.~Comp.~App.~Math.}}

\newcommand{\vecv}{\mathbf{v}}
\newcommand{\vecG}{\mathbf{G}}
\newcommand{\vecn}{\mathbf{n}}
\newcommand{\vecF}{\mathbf{F}}
\newcommand{\vecPi}{\mathbf{\Pi}}

\title[AMR Radiation Hydrodynamics] 
{Star Formation with Adaptive Mesh Refinement Radiation Hydrodynamics}

\author[Mark R.~Krumholz]   
{Mark R.~Krumholz$^1$}

\affiliation{$^1$Dept. of Astronomy \& Astrophysics, University of California, Santa Cruz, Interdisciplinary Sciences Building, Santa Cruz, CA 95064, USA \\ email: {\tt krumholz@ucolick.org}}

\pubyear{2010}
\volume{270}  
\pagerange{xxx}
\jname{Computational Star Formation}
\editors{J.~ Alves, B.~Elmegreen, J.~Girart, \& V.~Trimble, eds.}
\begin{document}

\maketitle

\begin{abstract}
I provide a pedagogic review of adaptive mesh refinement (AMR) radiation hydrodynamics (RHD) methods and codes used in simulations of star formation, at a level suitable for researchers who are not computational experts. I begin with a brief overview of the types of RHD processes that are most important to star formation, and then I formally introduce the equations of RHD and the approximations one uses to render them computationally tractable. I discuss strategies for solving these approximate equations on adaptive grids, with particular emphasis on identifying the main advantages and disadvantages of various approximations and numerical approaches. Finally, I conclude by discussing areas ripe for improvement.
\end{abstract}

\firstsection 
\section{Introduction}

While gravity is the dominant force in star formation, radiative processes are crucial as well. Most basically, radiation removes energy, allowing a collapsing cloud to maintain a nearly constant, low temperature as its density and gravitational binding energy rise by many orders of magnitude. Radiative cooling is what makes star formation a dynamic rather than a quasi-static process, a point understood quite early \citep[see the review by][]{hayashi66a}. At densities below $\sim 10^{4-5}$ H atoms cm$^{-3}$, cooling occurs primarily via collisionally-excited atomic or molecular line emission, while at higher densities dust and gas become thermally well-coupled via collisions, and thermal radiation by dust grains is the dominant cooling process \citep[e.g.][]{goldsmith01a}. At yet higher densities the coupled dust-gas fluid becomes optically thick to its own cooling radiation, causing the temperature to rise in a quasi-adiabatic fashion \citep{larson69}.

While one-dimensional simulations of this collapse can include quite sophisticated treatments of radiative transfer between gas parcels \citep[e.g.][]{masunaga98}, the computational difficulty of handling RHD in three dimensions led to a long period in which most 3D simulations of star formation simply dispensed with radiation entirely, choosing instead to represent the cooling processes either by a cooling function that removes energy at a rate dependent only on local gas properties, or by an even simpler equation of state that prescribes the gas temperature as a function of density. However, this approach has a major flaw: it does not allow energy exchange between parcels of gas, or between stars and diffuse gas. The last few years have seen a number of studies pointing out that this energy exchange is crucial to the dynamics of star formation, and that it can be dealt with only by RHD simulations.

The most important radiative transfer processes can be broken into three rough categories: thermal feedback, in which collapsing gas and stars heat the gas and thereby change its pressure; force feedback, in which radiation exerts forces on the gas that alter its motion; and chemical feedback, in which radiation changes the chemical state of the gas (e.g.\ by ionizing it), and this chemical change somehow affects the dynamics. Each of these processes has been the object of intense study in the last few years.

With regard to thermal feedback, \citet{larson05} pointed out that how gas fragments, and thus the stellar mass distribution that results from fragmentation, is extremely sensitive to how the temperature varies with density. \citet{krumholz06b} and \citet{krumholz08a} showed that one consequence of this sensitivity is that radiation produced by the first few stars to form in a given region can heat the gas around them, reducing the ability of that gas to fragment, favoring monolithic collapse to massive stars and suppressing formation of low mass objects. Subsequent numerical simulations \citep{krumholz07a, krumholz10a, bate09a, offner09a, urban10a} using a variety of methods have confirmed this effect. 

Force feedback becomes important in the context of massive stars, which produce such high luminosities that the radiative force they exert on dusty interstellar gas can exceed their gravitational force. A number of authors have argued based on one-dimensional models that this process sets an upper limit on stellar masses \citep{kahn74, yorke77, wolfire87}. More recent work in two (\citealt{nakano89, nakano95, jijina96,yorke99,yorke02}; also see the contribution by Kuiper et al.\ in these proceedings) and three \citep{krumholz07a, krumholz09c} dimensions instead suggests that force feedback does not limit stellar masses. Deciding the question requires radiation-hydrodynamic simulations.

The most important type of chemical feedback is photoionization, which raises the gas from $\sim 10$ K to $\sim 10^4$ K. This causes a number of important dynamic effects, including ejecting mass from star-forming clouds \citep{dale05, peters10a}, altering the magnetic field structure \citep{krumholz07f}, and driving turbulent motions \citep{gritschneder09a}. Radiation also drives other chemical processes, but these are usually not important for dynamics, and thus may be handled by post-processing simulations. In this review I limit myself to radiative effects that are dynamically important and must therefore be simulated in tandem with the hydrodynamic evolution of the system.

\section{The Equations of Radiation Hydrodynamics}

The equations of radiation hydrodynamics (RHD) in conservation form read \citep{mihalas99}
\begin{equation}
\label{eq:conslaws}
\frac{\partial}{\partial t} \left(
\begin{array}{c}
\rho \\
\rho\vecv \\
\rho e
\end{array}
\right)
+
\nabla\cdot
\left(
\begin{array}{c}
\rho\vecv \\
\rho \vecv:\vecv + P\\
(\rho e+P)\vecv
\end{array}
\right)
=
\left(
\begin{array}{c}
0 \\
\vecG \\
c G^0
\end{array}
\right),
\end{equation}
where $\rho$, $\vecv$, $e$, and $P$ are the gas density, velocity, specific energy, and pressure. The source term on the right hand side represents the rate at which radiation transfers momentum and energy to the gas. For simplicity I have omitted terms describing gravity and magnetic fields, which would appear as additional sources on the right hand side.

The rate at which radiation transfers energy and momentum to the gas is described by the radiation four-force vector $(G^0, \vecG)$,
\begin{equation}
\label{eq:forcevec}
c\left(
\begin{array}{c}
G^0 \\
\vecG
\end{array}
\right)
=
\int \, d\nu \int \, d\Omega \,\left[\kappa_\nu(\vecn) I_\nu - \eta_\nu(\vecn)\right]
\left(
\begin{array}{c}
1 \\
\vecn
\end{array}
\right)
\end{equation}
where $\vecn$ is a unit vector, $I_\nu$ is the radiation intensity at frequency $\nu$ in direction $\vecn$, and $\kappa_\nu$ and $\eta_\nu$ are the extinction coefficient and emissivity of the gas as a function of frequency $\nu$ and direction $\vecn$. Finally, the intensity is governed by the transfer equation,
\begin{equation}
\label{eq:transfer}
\frac{1}{c} \frac{\partial}{\partial t} I_\nu + \vecn \cdot \nabla I_\nu = \eta_\nu(\vecn) - \kappa_\nu(\vecn) I_\nu.
\end{equation}
For simplicity I have omitted terms describing scattering, which is generally not important for the radiation-hydrodynamics of star formation.

The equations of RHD are, like the ordinary equations of fluid dynamics, characterized by dimensionless numbers. The two most important ones for radiation-hydrodynmics are $\beta \sim v/c$, where $v$ is the characteristic value of $\vecv$, and $\tau \sim L/\lambda_p$, where $L$ is the size-scale of the system and $\lambda_p$ is the photon mean free path. The first ratio $\beta$ characterizes how relativistic the system is. In writing equation (\ref{eq:conslaws}) in non-relativistic form, we have already assumed that $\beta \ll 1$. The second, $\tau$, characterizes how optically thick it is. Systems with $\tau \ll 1$ are described as being in the streaming limit, and are characterized by weak matter-radiation interaction. Those with $\tau\gg 1$ are in the diffusion limit and have strong-matter radiation coupling. Both limits appears in star formation problems.

\section{Approximations and Solution Methods}

The full system of equations (\ref{eq:conslaws}) and (\ref{eq:transfer}) is seven-dimensional, with quantities varying in time, space (3 dimensions), direction (2 dimensions), and frequency. Unfortunately, this makes full numerical solution prohibitively expensive even on modern supercomputers. We are therefore reduced to approximations. There are two broad classes of approximation in common use in star formation simulations: moment methods and characteristic methods. A third category, Monte Carlo methods, has been used extensively in post-processing radiative transfer calculations, but has not been used extensively in radiation-hydrodynamic simulations. I will not discuss it in detail.

\subsection{Codes}

In the discussion that follows I introduce the most common methods used in AMR RHD methods, and in Table \ref{tab:codes} I summarize which method(s) are used in each code. The codes I include in the table are as follows: {\bf Orion} is the oldest and probably best-tested AMR RHD code, but is not publicly available, and currently lacks a magnetohydrodynamic (MHD) capability \citep{klein99, howell03, krumholz07b}. {\bf Ramses} is an AMR MHD code that was recently upgraded with a radiation solver, although the latter is not (as of this writing) publicly available \citep{fromang06, commercon10a}. {\bf Flash} is an open source AMR MHD code with extensive chemistry capabilities and several add-on modules that handle radiation in different ways \citep{fryxell00a, rijkhorst06a, peters10a}. {\bf Enzo} is an open source AMR HD code (an MHD version exists but is not public) with cosmology capabilities and two different radiation methods \citep{abel02, norman08a, reynolds09a}. {\bf Pluto} is an open source AMR MHD code with a newly-developed radiation solver \citep{mignone07a, kuiper10a}. There are also a number of fixed, nested grid codes in use, but I will not mention these by name.

\begin{table}
\begin{center}
\begin{tabular}{lccclcc}
\hline\hline\\[-6pt]
\multicolumn{3}{c}{Moment methods} 
& \hspace{0.4in} & \multicolumn{3}{c}{Characteristic methods}
\\ \hline
{\bf Code} & {\bf $\nu$ resolution} & {\bf Frame} 
& & {\bf Code} & {\bf $\nu$ resolution} & {\bf Ray scheme}
\\ \hline
Orion & MG, 2T & Mixed 
& & Orion & 1F & HEALpix 
\\
Ramses & 2T & Comoving 
& &
Flash & MG & Hybrid characteristics
\\
Enzo & 2T & Comoving 
& & Enzo & 1F & HEALpix
\\
Pluto & 1T & Comoving
& &
Pluto & MG & Sphere \\[2pt] \hline\hline
\end{tabular}
\end{center}
\caption{
\label{tab:codes}
Summary of Codes. See text for explanation of fields. All moment codes use first order closure (flux-limited diffusion). Enzo and Orion appear in both lists because they can operate in characteristic or moment mode. Pluto appears ion both lists because it uses a hybrid characteristic-moment method (see text for details). Orion has both a MG and a 2T moment method.
}
\end{table}

\subsection{Moment Methods}

\subsubsection{Basic Theory of Moment Methods}

The basic idea of a moment method is to take moments of the transfer equation, in exact analogy to the Chapman-Enskog procedure used to derive the equations of fluid dynamics from the kinetic theory of gases. To take the zeroth moment we integrate the transfer equation over all directions; to obtain the first moment we multiply both sides by the unit vector $\vecn$ and then integrate; for the second moment we multiply by the rank two tensor $\vecn:\vecn$ and integrate, and so forth.  As in the analogous fluid case, the procedure yields an exact solution if carried out to infinitely many orders, but one instead makes an approximation by truncating the procedure after finitely many orders. Usually ``finitely many" here means one or two, and the first two moments of equation (\ref{eq:transfer}) are
\begin{equation}
\frac{\partial}{\partial t}
\left(
\begin{array}{c}
E \\
\vecF/c^2
\end{array}
\right)
+ \nabla \cdot
\left(
\begin{array}{c}
\vecF \\
\vecPi
\end{array}
\right)
=
-\left(
\begin{array}{c}
c G^0 \\
\vecG
\end{array}
\right),
\label{eq:momenteqn}
\end{equation}
where
\begin{eqnarray}
E & = & \frac{1}{c} \int \, d\nu \int \, d\Omega \, I_\nu \\
\vecF & = & \int \, d\nu \int \, d\Omega \, I_\nu \vecn \\
\vecPi & = & \frac{1}{c} \int \, d\nu \int \,d\Omega \, I_\nu \vecn:\vecn
\end{eqnarray}
are the first three moments of the radiation intensity: the radiation energy density, radiation flux, and radiation pressure tensor. In these equations the direction-dependence in the transfer equation is removed and the dimensionality of the problem is lowered by two, but at the price of introducing the radiation pressure tensor, for which we do not have an equation (since we have not expanded the transfer equation to the next order) and must instead make an approximation.

Because moment methods necessarily involve smoothing the angular dependence of the radiation field, they are best suited to representing diffuse, smooth radiation fields. This makes them ideal for handling thermal and force feedback, which tend to be dominated by diffuse infrared light re-radiated by dust grains. They are less suited for chemical feedback, which is usually dominated by direct, highly beamed radiation from a handful of stellar sources. A further advantage of moment methods is that their computational cost is independent of the number of sources, and usually scales only as $N$ or $N\log N$, where $N$ is the number of cells. They also parallelize fairly easily.

\subsubsection{Types of Moment Methods}

Moment methods can be classified based on several design choices that are made in their construction, all of which involve some tradeoff between computational expense and accuracy. See Table \ref{tab:codes} for a list of the design choices made in each of the popular codes.

{\it Closure order.} The first choice is whether to retain both of the first two moment equations, or to retain only the zeroth moment and adopt a closure approximation for the radiation flux; all the moment methods currently used in practical star formation simulations make the latter choice, using a closure known as flux-limited diffusion (FLD). In a medium where the photon mean free path is small compared to the size of the system, Eddington showed that the radiation flux in the fluid rest frame approaches
\begin{equation}
\vecF = - \frac{c}{3\kappa_{\rm R}} \nabla E,
\end{equation}
where $\kappa_R = \int \, d\nu\, (\partial B_{\nu}/\partial T) / \int \,d\nu\, \kappa^{-1} (\partial B_{\nu}/\partial T)$ is the Rosseland mean opacity, $B_{\nu}(T)$ is the Planck function. This is known as the diffusion approximation. While it is quite good in optically thick media, it breaks down in optically thin regions because as $\kappa_{\rm R}\rightarrow 0$ the signal speed approaches infinity rather than properly limiting to $c$. The FLD approximation is an attempt to fix this problem by instead setting the radiation flux to
\begin{equation}
\vecF_0 = - \frac{c\lambda}{\kappa_{\rm R}} \nabla E_0,
\end{equation}
where $\lambda$ is a function that has the property that $\lambda\rightarrow 1/3$ in optically thick regions and $\lambda\rightarrow \kappa_{\rm R} E_0/\nabla E_0$ in optically thin regions, so that $|\vecF_0| \rightarrow cE_0$, its correct limiting value. Many choices for the function $\lambda$ are possible; the most common one in astrophysical applications is the \citet{levermore81} limiter, $\lambda(R) = R^{-1} (\mbox{coth}\, R - R^{-1})$, where $R=|\nabla E_0|/(\kappa_{\rm R} E_0)$. However, all of these limiters are of unknown accuracy in the intermediate optical depth regime, and all FLD methods suffer from the problem that they discard information about the directionality and momentum content of the radiation field. Higher order moment methods exist and can solve some of these problems (e.g.\ variable tensor Eddington factor methods, \citealt{hayes03}, and the M1 closure method, \citealt{gonzalez07a}), but none have yet proven cheap and robust enough for use in practical AMR star formation simulations.

{\it Frequency resolution.} The second choice in setting up an RHD moment method is the level of resolution in frequency. Most accurate is the multigroup (MG) method, in which one discretizes equation (\ref{eq:momenteqn}) in frequency, solving one version of the equation for each frequency bin. All the bins are coupled via the matter temperature, which affects the extinction and emissivity and thus the radiation four-force vector (equation \ref{eq:forcevec}). Next most accurate is the 2T method, in which one takes the radiation spectrum to be a Planck function characterized by a single temperature $T_{\rm rad}$ (thus removing a dimension from the problem), but allows $T_{\rm rad}$ to be different than the gas temperature $T_{\rm gas}$. Simplest of all is the 1T method, in which one assumes $T_{\rm rad} = T_{\rm gas}$, allowing a further simplification of the equations. As one might expect, this choice involves a tradeoff between accuracy and cost; the 1T method is cheapest but badly misestimates both the radiation spectrum and $T_{\rm gas}$ in the streaming regime. In comparison 2T methods still produce incorrect radiation spectra in the streaming regime, but make much smaller errors in the matter temperature. They are intermediate in cost. MG provides a good representation of both the spectrum and the matter temperature but is most expensive. 

{\it Choice of frame.} The third design choice is whether to formulate the equations in the comoving frame or using mixed frames. Extinction $\kappa$ and emissivity $\eta$ are simple, isotropic functions in the frame comoving with the fluid, but in other frames they contain complex directional dependence as a result of relativistic boosting. One might think that this effect is small in non-relativistic flows, but ignoring it is equivalent to neglecting the work done by radiation on the gas \citep{mihalas82}, which is obviously unacceptable if radiation forces are non-negligible. To avoid complex velocity dependence in $\eta$ and $\kappa$, it is highly desirable to write them in the comoving frame. The simplest choice after that is to write the radiation quantities ($E$, $\vecF$) in the comoving frame as well (e.g.\ see \citealt{mihalas82}), but this carries a significant cost. Since the comoving frame is non-inertial in any system with non-constant fluid velocity, the comoving radiation energy is not a conserved quantity, and thus comoving-frame codes cannot be exactly conservative. Moreover, every time the resolution changes and the velocity field is refined, the reference frame change as well, introducing errors in conservation that are likely to accumulate with increasing refinement. The alternative is a mixed-frame formulation in which one writes the radiation quantities in the inertial lab frame. Deriving this formulation is tricky, since one must carefully account for the Lorentz transformations between frames, but the resulting equations are explicitly conservative, and can be discretized in a manner that conserves energy to machine precision \citep{krumholz07b}.

\subsection{Characteristic Methods}

In characteristic methods one solves the transfer equation (\ref{eq:transfer}) directly, but only along selected rays. Given this solution, one can compute the radiation four-force vector (equation \ref{eq:forcevec}) directly. This provides much greater accuracy in computing the radiation intensity along those rays, but at the cost of neglecting all other rays. For this reason it is best used for chemical feedback, which tends to be dominated by the rays coming from a small number of sources. It is not well-suited to handling diffuse radiation fields, and simulations based on this technique are generally no better than non-radiative simulations when it comes to handling effects like fragmentation being altered by a diffuse IR radiation field. The cost of these methods scales as the number of cells times the number of sources, but with a coefficient that is generally smaller than for moment methods. Thus these methods are cheaper than moment methods when the number of sources and rays is small, but become impractically expensive if the number of sources is even a small fraction of the number of cells. As with moment methods, there are several design choices to be made in setting up a characteristic method.

{\it Frequency resolution.} As in the moment method case, one can choose either to retain the frequency-dependence in the transfer equation by using multiple frequency groups (the MG method), or one can integrate over frequency. In this case one does not generally assign a temperature to the radiation; instead, since for chemical feedback one cares about photons only above a certain energy, the usual approximation is to adopt an average photon energy. This is the single-frequency (1F) approximation.

{\it Ray-drawing scheme.} There are also a number of schemes for drawing rays from the point source(s). The simplest is the spherical ray technique: one uses spherical coordinates, requires that the source be at the origin, and simply draws rays aligned with the computational grid. This parallelizes almost perfectly and is very simple to code, but is obviously unsuited to any problem with multiple stellar sources or where the locations at which stars form is not known {\it a priori}. The other two schemes in wide use are HEALPix-based adaptive trees \citep{abel02} and the hybrid characteristics \citep{rijkhorst06a}, both of which operate on top of an underlying Cartesian grid. The HEALPix scheme is based on the Hierarchical Equal Area isoLatitude Pixelization of the sphere, which divides the sphere into a equal area pixels that can be subdivided indefinitely, allowing the angular resolution to adapt to match that of the underlying grid. The hybrid characteristics scheme works by combining a method of short characteristics within individual AMR grids with a method of long characteristics between grids. Neither suffer from the limitations of the spherical ray method.

\subsection{Hybrid Methods}

Hybrid methods attempt to combine the best features of characteristic and moment methods. Characteristics work well for the highly beamed radiation coming directly from a star or some other point source. However, once this radiation is absorbed, following its re-emission with characteristics becomes unreasonably expensive, and a moment method is a far better choice. The underlying idea of a hybrid method, first used in multidimensional simulations by \citet{murray94a} and recently implemented in the Pluto code \citep{kuiper10a}, is to use characteristics for the ``first absorption", then switch to a moment method. One does this by performing a characteristic trace from the star or stars to determine the rate of energy and momentum input into each cell by direct radiation. One adds this as a source term on the right hand side of the moment equation (\ref{eq:momenteqn}), then solves as one would in a pure moment method.  The computationally-cheap characteristic step can use MG, while the more expensive moment solve uses the 1T or 2T approximation. Since the first absorption is generally the part of the problem where the frequency dependence is most important, this method achieves some of the accuracy of a fully frequency-dependent calculation at significantly lower cost.

\section{Future Directions}

I close this review with a brief discussion of possible future directions for AMR RHD in the star formation context. One such likely direction comes from combining realistic treatments of thermal, force, and chemical feedback into a single simulation, rather than treating only one or two of them at a time as in current simulations. Since the first two effects are most easily handled by moment methods and the third by characteristic methods, hybrid techniques are the natural solution. To be competitive in handling thermal and force feedback, however, these methods will have to be linked to more advanced moment methods than has been attempted before. Nonetheless, a hybrid method combining frequency-dependent characteristic tracing with a 2T, mixed frame FLD method is a straightforward extension of current techniques, and seems a logical place to start.

Further in the future, second order moment methods are likely to become important. Developing them to the point where they can run reliably in a parallel environment on adaptive grids will be a major algorithmic undertaking, one that is likely to require the assistance of computer scientists and applied mathematicians in addition to astronomers. If successful, these methods will be able to handle both beamed and diffuse radiation fields within a single framework, and will remove many of the limitations of the FLD approximation. Monte Carlo methods are another candidate to replace the current dominance of FLD and characteristic methods. They are very computationally costly, since many photons are required to suppress Poisson noise, but they have the advantage of near perfect parallelization. Since the future of supercomputing seems to be heading toward the development of ever-larger numbers of processors, each of which is no faster than the processors of the previous generation, this may prove to be a decisive advantage.

Finally, a caveat: for all the limitations of our current RHD methods, it is not entirely obvious that our numerical techniques are the limiting factor in the accuracy of our simulations. The main agent for coupling radiation and gas at the high densities where stars form is dust. Our knowledge of dust grain properties, such as sublimation temperatures and grain size distributions in dense environments, is quite limited, and appears likely to advance less quickly than either computer power or algorithmic technique. It may be the case in the future that our knowledge of dust becomes the limiting factor in the accuracy of RHD simulations.

\begin{acknowledgments}
I thank Richard Klein, Chris McKee, and Loius Howell for helpful discussions. I acknowledge support from: an Alfred P.\ Sloan Fellowship; NASA through ATFP grant NNX09AK31G and through the Spitzer Theoretical Research Program, through a contract issued by the JPL; and the NSF through grant AST-0807739.
\end{acknowledgments}

\end{document}